\documentclass[3p,twocolumn]{elsarticle}
\usepackage{amsmath,amssymb,graphicx,hyperref}
\usepackage[capitalise]{cleveref}
\hypersetup{colorlinks=true}

\begin{document}
	\begin{frontmatter}
		\title{Theory of the dipole-exchange spin wave spectrum in ferromagnetic films with in-plane magnetization revisited}
		
		\author[1]{J. S. Harms\corref{cor1}}
		\ead{J.S.Harms@uu.nl}
		
		\author[1,2]{R. A. Duine}
		\ead{R.A.Duine@uu.nl}
		\address[1]{Institute for Theoretical Physics, Utrecht University, 3584CC Utrecht, The Netherlands}
		\address[2]{Department of Applied Physics, Eindhoven University of Technology, P.O. Box 513, 5600 MB Eindhoven, The Netherlands}
		
		\cortext[cor1]{Corresponding author}
	
	\begin{abstract}
	We present a refinement of the widely accepted spin-wave spectrum that Kalinikos and Slavin~\cite{kalinikos_spectrum_1981,kalinikos_theory_1986} computed for magnetic films with an in-plane magnetization.
	The spin wave spectrum that follows from the diagonal approximation in this theory becomes inaccurate for relatively thick films,
	as has already been noted by~Kreisel \textit{et al.}~\cite{kreisel_microscopic_2009}.
	Rather than solving an integrodifferential equation which follows from the magnetostatic Green's function, as done by Kalinikos and Slavin~\cite{kalinikos_spectrum_1981,kalinikos_theory_1986}, we impose the exchange and magnetostatic boundary conditions on bulk spin-wave solutions.
	This boundary problem has an accurate analytical solution which is quantitatively different from the commonly used diagonal theory~\cite{kalinikos_spectrum_1981,kalinikos_theory_1986} for magnetic films.
	\end{abstract}
	\end{frontmatter}
	\section{Introduction}
	Dipole-exchange spin waves propagating in in-plane magnetized magnetic thin films have attracted lot of attention in recent years, due to their potential applications in magnonic devices~\cite{chumak_magnon_2015}. Of special interest is the case in which spin waves travel perpendicular to the external magnetic field -- in which case the spin wave velocity is the largest.
	As noted by~Kreisel~\textit{et al.}~\cite{kreisel_microscopic_2009} the spin wave spectrum that follows from the diagonal approximation in the commonly used theory~\cite{kalinikos_spectrum_1981,kalinikos_theory_1986} is inaccurate in this case for relatively thick films.
	The inaccuracy stems from the diagonal approximation, and disappears when solving the system numerically with interband interactions.
	This approach on the other hand is not feasible for analytic approximations. 
	
	In this article, we present an alternative analytic derivation of the dipole-exchange spin wave spectrum for this scenario.
	Rather than solving an integrodifferential equation following from the magnetostatic Green's functions as done by Kalinikos and Slavin~\cite{kalinikos_spectrum_1981,kalinikos_theory_1986},
	we use an approach resembling that of~\citet{wolfram_surface_1972} which previously had no analytical solution.
	A similar approach has been used by~\citet{sonin_is_2017} to derive the spectrum of spin waves propagating parallel to an in-plane magnetic field for sufficiently large wave numbers. 
	Kalinikos and Slavin~\cite{kalinikos_spectrum_1981,kalinikos_theory_1986} approximately solved the integrodifferential equations by assuming a superposition of magnetization profiles which satisfy the exchange boundary conditions but do not satisfy the bulk equations of motion. 
	Here, however we impose both the exchange and magnetostatic boundary conditions on bulk spin wave solutions.
	This boundary problem turns out to have an accurate analytical solution -- compared with the numerical spectrum -- and is quantitatively different from the commonly used diagonal spin wave theory~\cite{kalinikos_spectrum_1981,kalinikos_theory_1986} for relatively thick films.
	\begin{figure}
		\includegraphics[width=\columnwidth]{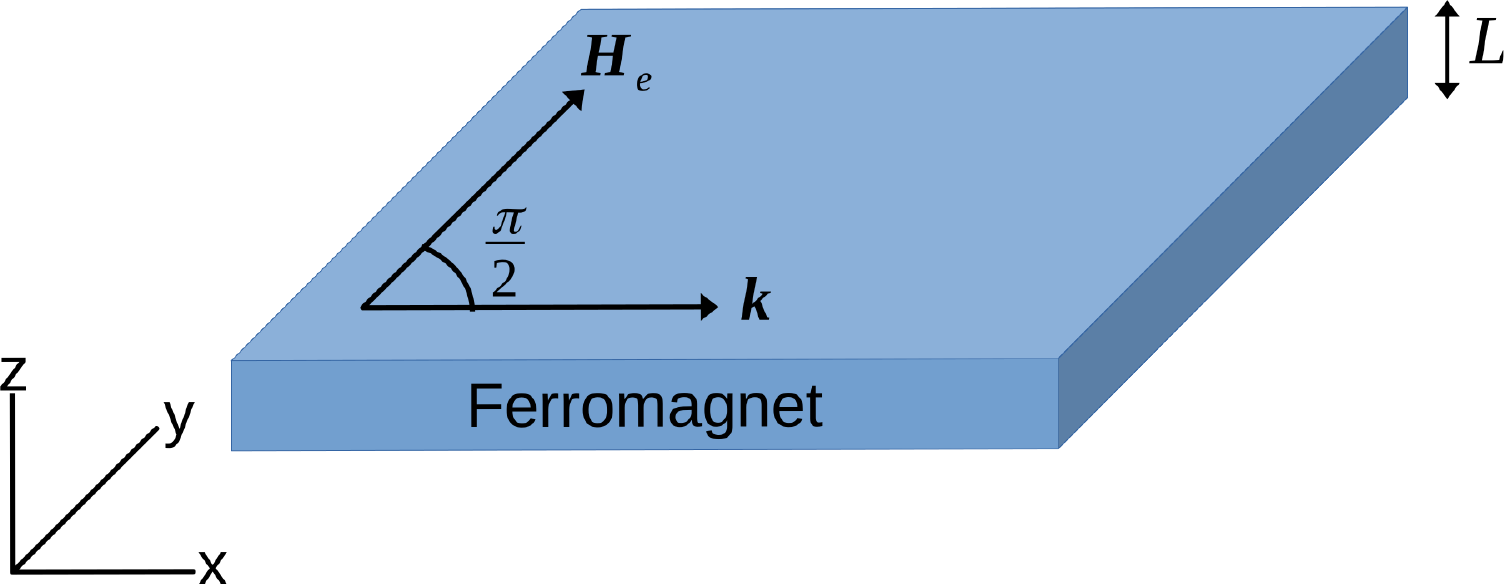}
		\caption{Sketch of the set-up. We consider a ferromagnetic thin film of thickness $ L $ with the equilibrium magnetization pointing in the $ y $ direction.}
		\label{fig:set-up}
	\end{figure}	
	\section{Thin-film ferromagnet}
	\subsection{Model and set-up}
	We consider the set-up in~\cref{fig:set-up} of a ferromagnetic thin film of thickness $ L $ subject to an in-plane external magnetic field $ \mathbf{H}_\mathrm{e} $.
	We chose the $ x-y $ axes to correspond to the in-plane directions, with the external magnetic field $ \mathbf{H}_{\mathrm{e}}=H_\mathrm{e}\hat{y} $ pointing in the $ y $ direction.
	Furthermore, the $ z $ axis corresponds to the out of plane direction where the surfaces of the thin film are located at $ z=\pm L/2 $.
	For temperatures below the Curie temperature, amplitude fluctuations in the magnetization are negligible.
	Hence, the dynamics of the magnetization direction $ \mathbf{n}=\mathbf{M}/M_s $ is described be the Landau-Lifshitz equation (LL), and the Maxwell equations in the magnetostatic limit -- accounting for dipole-dipole interactions.
	The LL equation is given by
	\begin{equation}\label{eq:LL_equation}
		\partial_t\mathbf{n}
		=
		-\gamma\mathbf{n}\times\mathbf{H}_{\mathrm{eff}},
	\end{equation}
	which describes precession of the magnetization direction around the effective field $ \mathbf{H}_{\mathrm{eff}}=-\delta E/\delta(M_s\mathbf{n}) $.
	Here, we consider the magnetic energy functional $ E[\mathbf{n}] $ of the form
	\begin{equation}\label{eq:magnetic-energy-functional}
		E[\mathbf{n}]
		=
		M_s\int dV
		\left[
		-\frac{1}{2}J\mathbf{n}\cdot\nabla^2\mathbf{n}
		-\mu_0 \mathbf{H}\cdot\mathbf{n}
		\right].
	\end{equation}
	In the above $ J $ is the spin stiffness and $ \mathbf{H}=\mathbf{H}_\mathrm{e}+\mathbf{H}_{\mathrm{D}} $ is the magnetic field strength, where $ \mathbf{H}_\mathrm{D} $ is the demagnetization field originating from dipole-dipole interactions.
	Furthermore, the magnetostatic Maxwell equations~\cite{jackson_classical_1998} -- accounting for dipole-dipole interactions -- are given by
	\begin{equation}\label{eq:magnetostatic-Maxwell-equations}
		\nabla\times\mathbf{H}=0,\space\nabla\cdot\mathbf{B}=0,
	\end{equation}
	with $ \mathbf{B}=\mu_0(\mathbf{H}+\mathbf{M}) $ the total magnetic field.
	The boundary conditions require the normal component of $ \mathbf{B} $ and the tangential components of $ \mathbf{H} $ to be continuous at the thin film surfaces.
	
	In equilibrium the LL equation requires the equilibrium magnetization $\mathbf{M}_{\mathrm{eq}}$ and the effective magnetic field strength $ \mathbf{H}_{\mathrm{eff}} $ to be parallel $ \mathbf{M}_\mathrm{eq}\parallel\mathbf{H}_{\mathrm{eff}} $.
	In this case the internal magnetic field strength $ \mathbf{H}_{\mathrm{eq}}=\mathbf{H}_\mathrm{e}+\mathbf{H}_\mathrm{D} $ has a contribution from the external magnetic field $ \mathbf{H}_\mathrm{e} $ and the demagnetization field $ \mathbf{H}_\mathrm{D}=-\hat{z} M_z $, originating from the magnetostatic boundary conditions. For an external magnetic field pointing in the $ y $ direction, as discussed in this article, the uniform equilibrium magnetization $ \mathbf{M}_{\mathrm{eq}} $ should also point in the $ y $ direction.

	Dipole-exchange spin-wave modes are generated by dynamical fluctuations of the magnetization and the demagnetizing field around the magnetostatic equilibrium 
	\begin{equation}
		\mathbf{M}=\mathbf{M}_\mathrm{eq}+\mathbf{m}(t),\;\mathbf{H}=\mathbf{H}_\mathrm{eq}+\mathbf{h}_\mathrm{D}(t),
	\end{equation}
	where $ \mathbf{m} $ is perpendicular to $ \mathbf{M}_\mathrm{eq} $ up to linear order in $ m_x $ and $ m_z $, lying in the $ x-z $ plane.
	The latter is a consequence of the magnitude of the magnetization being constant $ |\mathbf{M}|=M_s $.
	Since the magnetostatic Maxwell equations~(\ref{eq:magnetostatic-Maxwell-equations}) are linear we require
	$
		\nabla\times\mathbf{h}_\mathrm{D}=0,\;
		\nabla\cdot\mathbf{b}=0,
	$
	with $ \mathbf{b}=\mu_0\left(\mathbf{h}_\mathrm{D}+\mathbf{m}\right) $. 
	Using that the dynamic demagnetizing field is has vanishing curl, we express the dynamic demagnetization field in terms of a scalar potential $ \mathbf{h}_\mathrm{D}=\nabla\Phi_\mathrm{D} $.
	Hence, the magnetostatic Maxwell equations become
	$
		\nabla^2\Phi_\mathbf{D}=-\nabla\cdot\mathbf{m},
	$
	where the magnetization outside the thin film vanishes.
	
		The Landau-Lifschitz and magnetostatic Maxwell equations may be rewritten by means of $ \mathbf{n}\simeq\hat{z}\,\sqrt{2}\mathrm{Re}[\Psi]-\hat{x}\,\sqrt{2}\mathrm{Im}[\Psi]+ \hat{y}\,\left(1-\left|\Psi\right|^2\right)$, with $ \Psi=(1/\sqrt{2})\left(\hat{z}-i\hat{x}\right)\cdot\mathbf{n} $. 
	Consequently the linearised LL and magnetostatic Maxwell equations become
	\begin{subequations}\label{eq:linearised_equations_of_motion}
		\begin{align}
			\label{eq:linearised_LLG}
			\begin{split}
				\hat{\Omega}\Psi=&-\left(\Omega_H-\Lambda^2\nabla^2\right)\Psi
				+ \frac{(\partial_{z}-i\partial_{x})}{\sqrt{2}M_s}\Phi_{\mathrm{D}},
			\end{split}
			\\
			\label{eq:linearised_Maxwell}	
			\frac{\nabla^2\Phi_{\mathrm{D}}}{M_s^2}=&-\frac{(\partial_{z}+i\partial_{x})}{\sqrt{2}M_s}\Psi-\frac{(\partial_{z}-i\partial_{x})}{\sqrt{2}M_s}\Psi^*.
		\end{align}
	\end{subequations}
	Additionally, the exchange boundary conditions for thin films~\cite{soohoo_general_1963} in the absence of surface anisotropy require
	\begin{equation}\label{eq:exchange-boundary-conditions}
	\pm\partial_z\Psi\big\rvert_{z=\pm L/2}=0.
	\end{equation}
	In the above, we defined the dimensionless magnetic field $ \Omega_H=\mu_0H_e/\mu_0M_s, $
	exchange length $ \Lambda=\sqrt{J/\mu_0M_s} $
	and the dimensionless frequency operator
	$
	\hat{\Omega}=i\partial_t/\gamma\mu_0M_s.
	$
	\subsection{Bulk dipole-exchange spin-waves and it's boundary conditions}
	Using the Bogoliubov ansatz, we write
	$
	\Psi(\mathbf{x},t)=u(\mathbf{x})e^{-i\omega t}+v^*(\mathbf{x})e^{i\omega t}
	~\text{and}~
	\Phi_\mathrm{D}(\mathbf{x},t)=w(\mathbf{x})e^{-i\omega t}+w^*(\mathbf{x})e^{i\omega t},
	$
	where
	$ 
	\begin{pmatrix}
		u(\mathbf{x}), &
		v(\mathbf{x}), &
		w(\mathbf{x})
	\end{pmatrix}
	\propto
	e^{i\mathbf{k}\cdot\mathbf{r}_\parallel}
	\begin{pmatrix}
		u(\mathbf{k},z), &
		v(\mathbf{k},z), &
		w(\mathbf{k},z)
	\end{pmatrix},
	$
	with
	$
	\mathbf{k}
	=
	\begin{pmatrix}
		k_x&k_y
	\end{pmatrix} $.
	In these coordinates the linearised LL and magnetostatic Maxwell equation~\cref{eq:linearised_equations_of_motion} become
	$
		\mathcal{G}\cdot	
		\begin{pmatrix}
			u(\mathbf{k},z)&
			v(\mathbf{k},z)&
			w(\mathbf{k},z)
		\end{pmatrix}
		=0
	$
	with
	\begin{equation}\label{eq:bulk_equations_of_motion}
		\begin{aligned}
		\mathcal{G}=
		\begin{pmatrix}
			-\sqrt{2}M_sF & 0 & (\partial_z+k_x)
			\\
			0 & -\sqrt{2}M_sF^* & (\partial_z-k_x)
		\\
			\frac{M_s}{\sqrt{2}}(\partial_z-k_x) & \frac{M_s}{\sqrt{2}}(\partial_z+k_x) & 
			(\partial_z^2-k^2)
		\end{pmatrix}.
		\end{aligned}
	\end{equation}
	and $ \hat{F}=\Omega+\Omega_h+\Lambda^2(k^2-\partial_z^2) $ , $ \hat{F}^*=-\Omega+\Omega_h+\Lambda^2(k^2-\partial_z^2) $ the dimesionless LL spin-wave operators and $ \Omega=\omega/\gamma\mu_0M_s $.
	The above bulk equation of motion gives rise to a sixth order homogeneous differential equation in position space, which is cubic with respect to $ \partial_z^2 $. For spin waves travelling in the $ x $ direction, perpendicular to the external magnetic field, the general solution of~\cref{eq:bulk_equations_of_motion} is given by the linear combination of plane waves
	\begin{equation}\label{eq:magnetostatic_bluk_ansatz}
		\begin{pmatrix}
			u(\mathbf{x})\\
			v(\mathbf{x})\\
			w(\mathbf{x})
		\end{pmatrix}
		=
		\sum_{l=1}^6
		C_{k_l}
		\begin{pmatrix}
			F_l^*(k_l+k_x)/\sqrt{2}M_s\\
			F_l(k_l-k_x)/\sqrt{2}M_s\\
			F_l^*F_l
		\end{pmatrix}
		e^{k_l z+ik_x x},
	\end{equation}
 	where the wave numbers $ k_l $ satisfy the bulk equations of motion which follow from setting the determinant of~\cref{eq:magnetostatic_bluk_ansatz} to zero
 	\begin{equation}\label{eq:bulk_equation_of_motion_no_boundary_conditions}
 		F_l^*F_l
 		(k^2-k_l^2)
 		+
 		(1/2)(F_l^*+F_l)
 		(k_x^2-k_l^2)=0,
 	\end{equation}
 	which is explicitly written as
 	\begin{equation}\label{eq:bulk-equation-of-motion-rewritten}
 		(k^2-k_l^2)
 		\Big\{
 		\big[\Omega_h+1/2
 		+
 		\Lambda^2
 		(k^2-k_l^2)\big]^2
 		-1/4
 		-\Omega^2
 		\Big\}
 		=0.
 	\end{equation}
	We find that the bulk equation of motion in~\cref{eq:bulk_equation_of_motion_no_boundary_conditions} gives rise to one volume mode $ k_z=\pm i q $ and two real surface modes $ k_z=\pm k_{1,2} $, with $ \{q,k_1,k_2\} $ real and positive.  Furthermore, the bulk equation of motion~\cref{eq:bulk-equation-of-motion-rewritten} may be rewritten in the dispersive form
	\begin{equation}\label{eq:dispersion_in_terms_of_volume_mode_q}
		\Omega^2
		=
		\left[\Omega_H+1/2+\Lambda^2k^2+\Lambda^2q^2\right]^2-1/4,
	\end{equation}
	where the precise for of the volume mode $ q $ follows from the boundary conditions on the system. From~\cref{eq:bulk_equation_of_motion_no_boundary_conditions,eq:dispersion_in_terms_of_volume_mode_q} we find that the remaining two surface modes $ k_{1,2} $ may be expressed as
	\begin{subequations}\label{eq:spin_wave_bulk_surface_modes}
		\begin{align}
			k_1^2&=k_x^2,\\
			k_2^2&=k_x^2+\Lambda^{-2}\left[2\Omega_H+1+\Lambda^2k^2+\Lambda^2q^2\right],
	\end{align}
	\end{subequations}
	where the mode with wave length $ k_1 $ is referred to as the Damon-Eshbach (DE) mode~\cite{damon_magnetostatic_1961}.
	
	The exchange boundary conditions in~\cref{eq:exchange-boundary-conditions} evaluated for the spin-wave modes in~\cref{eq:magnetostatic_bluk_ansatz} give
	\begin{subequations}\label{eq:exchange-boundary-conditions-plane-wave-ansatz}
		\begin{align}
			\sum_{l=1}^6C_{k_l} k_lF^*_l(k_l+k_x)e^{\pm k_l L/2}\big\rvert_{z=\pm L/2}&=0,
			\\
			\sum_{l=1}^6C_{k_l} k_lF_l(k_l-k_x)
			e^{\pm k_l L/2}\big\rvert_{z=\pm L/2}&=0,
		\end{align}
	\end{subequations}
	with $ k_l\in\left\{\pm k_1,\pm k_2,\pm iq\right\}$ as defined in~\cref{eq:bulk_equation_of_motion_no_boundary_conditions,eq:spin_wave_bulk_surface_modes}.
	The magnetostatic boundary conditions on the other hand require a bit more work.
	We start by noting that the magnetization vanishes $ \Psi=0 $ outside the magnetic thin film $ (z<-L/2~\text{and}~L/2<z) $. Hence, the magnetostatic Maxwell equations~(\ref{eq:magnetostatic-Maxwell-equations}) outside the thin film give
	$
	\nabla^2w(\mathbf{x})=(-k^2+\partial_z^2)w(\mathbf{k},z)e^{i\mathbf{k}\cdot\mathbf{r}_\parallel}=0.
	$
	The asymptotically bound solutions of outside the magnet are thus given by 
	\begin{equation}\label{eq:scalar-potential-outside-thin-film}
		w(\mathbf{k},z)
		\propto\left\{
		\begin{array}{r l}
			e^{-k z}, & z>L/2,\\
			e^{k z}, & z<-L/2.
		\end{array}
		\right.
	\end{equation}
	Since the tangential components of $ \mathbf{h}_\mathrm{D} $ are continuous across the thin film surfaces, 
	the scalar field $ w(\mathbf{k},z) $ should also be continuous across the thin film surface. Furthermore, continuity of the normal component of $ \mathbf{b} $ in combination with~\cref{eq:scalar-potential-outside-thin-film} gives the effective magnetostatic boundary condition
	\begin{equation}\label{eq:magnetostatic_boundary_condition}
		(\pm k+\partial_z)w(\mathbf{k},z)
		+
		\frac{M_s}{\sqrt{2}}[u(\mathbf{k},z)+v(\mathbf{k},z)]\rvert_{z=\pm L/2}
		=0.
	\end{equation} 
	Evaluated for the spin-wave modes in~\cref{eq:magnetostatic_bluk_ansatz} the above effective magnetostatic boundary condition~(\ref{eq:magnetostatic_boundary_condition}) gives
	\begin{subequations}\label{eq:magnetostatic-boundary-conditions-spin-wave}
		\begin{align}
			&\sum_\pm C_{\pm k_1}\big[\big(2F^*_1F_1+F^*_1\big)\delta_\pm-F_1\delta_\mp\big]e^{\pm k_1 L/2}
			\\\nonumber&
			-\sum_{l}C_{k_l}F^*_le^{k_l L/2}\rvert_{z=L/2}=0,
			\\
			&\sum_\pm C_{\pm k_1}\big[(2F^*_1F_1+F_1)\delta_\mp-F^*_1\delta_\pm\big]e^{\mp k_1 L/2}
			\\\nonumber&
			-\sum_{l}C_{k_l}F_le^{- k_l L/2}\rvert_{z=-L/2}=0,
		\end{align}
	\end{subequations}
	with 
	$
	\delta_\pm
	\equiv\bigg\{
	\begin{matrix}
		1& +\\
		0& -
	\end{matrix},
	$ 
	and
	$ k_l\in\left\{\pm k_2,\pm iq\right\}$ as defined in~\cref{eq:bulk_equation_of_motion_no_boundary_conditions,eq:spin_wave_bulk_surface_modes}.
	Note here that we used the bulk equation of motion in~\cref{eq:bulk_equation_of_motion_no_boundary_conditions} to simplify the above boundary conditions.
	\section{Dipole-exchange dispersion relation}
	\subsection{General derivation}
	For notational simplicity we introduce the dimensionless wavenumbers $ \Lambda k\rightarrow k $, $ \Lambda q\rightarrow q $ and the dimensionless thickness $ L/\Lambda\rightarrow L $.
	\subsubsection{Effective boundary conditions for spin waves}
	We start this section with by noting that that~\cref{eq:spin_wave_bulk_surface_modes} yields $ k_2\gg k\equiv k_x $, $ e^{k_2L/2}\gg e^{-k_2L/2} $ and $ |F_2^*|\gg|F_2| $.
	This allows us to approximate the exchange boundary conditions in~\cref{eq:exchange-boundary-conditions-plane-wave-ansatz} by
	\begin{subequations}\label{eq:exchange_boundary_conditions_analytic_approximation}
	\begin{align}
		&a_+ F_{k}^*k^2 e^{kL/2}+ b_+=\\\nonumber
		&cF_q^*\big(q^2\cos[(q+\delta)L/2]
		+kq\sin[(q+\delta)L/2]\big),
		\\
		&a_+ F_{k}^*k^2 e^{-kL/2}+ b_-=\\\nonumber
		&cF_q^*\left(q^2\cos[(q-\delta)L/2]-kq\sin[(q-\delta)L/2]\right),
		\\
		&a_- F_kk^2e^{-kL/2}=\\\nonumber
		&cF_q\left(q^2\cos[(q+\delta)L/2]-kq\sin[(q+\delta)L/2]\right),
		\\
		&a_-F_kk^2e^{kL/2}=\\\nonumber
		&cF_q\left(q^2\cos[(q-\delta)L/2]+kq\sin[(q-\delta)L/2]\right).
	\end{align}
	\end{subequations}
	where $ a_\pm=C_{\pm k}, $ $ b_\pm\sim C_{\pm k_2} F_2^*k_2^2/2 $ and $ e^{\pm i\delta L/2}c=C_{\pm q} $. Note that $ \delta $ can in principle be a complex number.
	For future convenience we rewrite the above exchange boundary conditions to
	\begin{subequations}\label{eq:exchange_boundary_conditions_rewritten_final_form_appendix}
	\begin{align}
		\bar{b}_++a_- F_kk^2e^{-kL/2}&=2cF_qq^2\cos[(q+\delta)L/2],
		\\
		\bar{b}_+-a_- F_kk^2e^{-kL/2}&=2cF_qkq\sin[(q+\delta)L/2],
		\\
		\bar{b}_-+a_- F_kk^2e^{kL/2}&=2cF_qq^2\cos[(q-\delta)L/2],
		\\
		-\bar{b}_- +a_- F_kk^2e^{kL/2}&=2cF_qkq\sin[(q-\delta)L/2].
	\end{align}
	\end{subequations}
	where $ \bar{b}_+=(F_q/F_q^*)b_++ a_+ (F_q/F_q^*)F_{k}^*k^2 e^{kL/2} $ and $ \bar{b}_{-}=(F_q/F_q^*)b_-+a_+ (F_q/F_q^*)F_{k}^*k^2 e^{-kL/2} $ are free parameters since  $ k_2\gg k $ implies that $ b_+ $ and $ b_- $ are -- to good approximation -- not restricted by the magnetostatic boundary conditions.
	From here we find that the contributions of $ q $ and $ \delta $ can be separated by making use of the trigonometric identities
	\[
	\left\{
	\begin{aligned}
	\cos[(q\pm\delta)L/2]&=\cos(qL/2)\cos(\delta L/2)\\&\mp\sin(qL/2)\sin(\delta L/2),\\
	\sin[(q\pm\delta)L/2]&=\sin(qL/2)\cos(\delta L/2)\\&\pm\cos(qL/2)\sin(\delta L/2).
	\end{aligned}
	\right.
	\]
	The above trigonometric identities allow us to express $ \bar{b}_+ $ and $ \bar{b}_- $ in~\cref{eq:exchange_boundary_conditions_rewritten_final_form_appendix}  in terms of the variables $ a_- $, $ k $ and $ q $, which gives
	\begin{subequations}\label{eq:exact_expression_bpm_appendix}
	\begin{align}
			\bar{b}_{+}
			&=
			\frac{a_{-}F_kk^2}{k^2+q^2}
			\bigg[(q^2-k^2)e^{-kL/2}
			\\\nonumber&+2kq\left(\csc(qL)e^{kL/2}
			-\cot(qL)e^{-kL/2}\right)\bigg],\\
			\bar{b}_{-}
			&=
			\frac{a_{-}F_kk^2}{k^2+q^2}
			\bigg[(q^2-k^2)e^{kL/2}
			\\\nonumber&
			+2kq\left(\cot(qL)e^{kL/2}-\csc(qL)e^{-kL/2}\right)\bigg].
	\end{align}
	\end{subequations}
	Hence,~\cref{eq:exchange_boundary_conditions_rewritten_final_form_appendix,eq:exact_expression_bpm_appendix} allow us to express $ 2cF_q^*F_qq^2\cos[(q\pm\delta)L/2] $ in terms of the variables $ a_- $, $ k $ and $ q $, which to leading order in exponential functions is given by
	\begin{subequations}\label{eq:exchange-boundary-conditions-constraint}
	\begin{align}\label{eq:exchange-boundary-conditions-constraint-a}
		cF_q\cos[(q+\delta)L/2] 
		&=
		\frac{a_{-}F_kk^2e^{kL/2}}{k^2+q^2}
		\\\nonumber&
		\times\left(e^{-kL}+kL\csc(qL)/qL\right),
		\\
		cF_q^*\cos[(q-\delta)L/2] 
		&=
		\frac{a_{-}F_kk^2e^{kL/2}}{k^2+q^2}\frac{F_q^*}{F_q}
		\\\nonumber&\times\left(1+kL\cot(qL)/qL\right).
	\end{align}
	\end{subequations}
	So far we have used the exchange boundary conditions~\cref{eq:exchange_boundary_conditions_analytic_approximation} to express $ 2cF_q^*F_qq^2\cos[(q\pm\delta)L/2] $ in terms of the variables $ a_- $, $ k $ and $ q $. From here, we impose the magnetostatic boundary conditions to find a closed expression for $ q $ satisfying all boundary conditions.
	
	The remaining magnetostatic boundary conditions~(\ref{eq:magnetostatic-boundary-conditions-spin-wave}), for $ k_2\gg k_x $, $ e^{k_2L/2}\gg e^{-k_2L/2} $ and $ |F_2^*|\gg|F_2| $, are well approximated by
	\begin{subequations}\label{eq:magnetostatic_maxwell_equations_appendix}
		\begin{align}
			a_+(2F_kF_k^*+F_k^*)e^{kL/2}-a_-F_ke^{-kL/2}=&\\\nonumber
			2cF_q\cos[(q+\delta)L/2]&,
			\\
			a_+F_k^*e^{-kL/2}-a_-(2F_kF_k^*+F_k)e^{kL/2}=&\\\nonumber
			-2cF_q^*\cos[(q-\delta)L/2]&.
		\end{align}
	\end{subequations}
	\subsubsection{Dipole-exchange spin-wave modes}
\begin{figure}
		\includegraphics[width=\columnwidth]{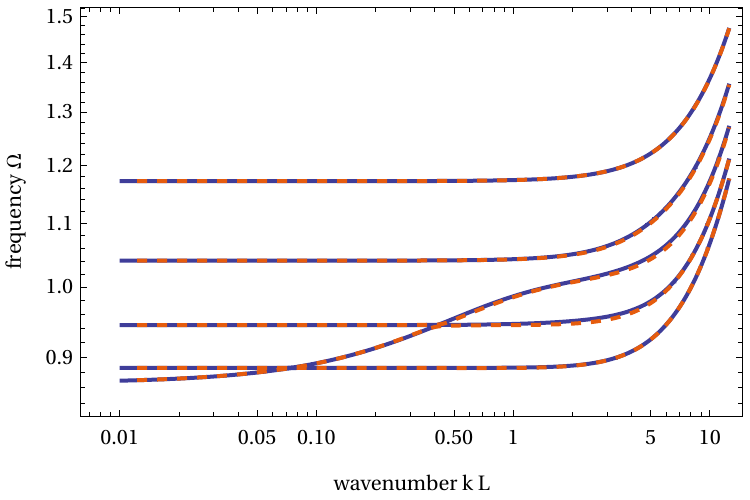}
		\caption{Dipole-exchange spin-wave dispersion relation for $ \Omega_H=1/2 $ and $ L=24 $. The dashed lines correspond to the analytic dispersion in~\cref{eq:spin-wave-dispersion-relation}, while the solid lines correspond to the full numeric solution of~\cref{eq:exchange-boundary-conditions-plane-wave-ansatz,eq:magnetostatic-boundary-conditions-spin-wave}.}
		\label{fig:spin-wave-dispersion-L-24}
	\end{figure}

	The above magnetostatic boundary conditions together with the effective exchange boundary conditions in~\cref{eq:exchange-boundary-conditions-constraint} give two linear homogeneous equations in $ a_+ $ and $ a_- $.
	Hence, we have spin-wave solutions when the determinant of this square matrix vanishes.
	At leading order in exponential functions of the trigonometric contribution we find this to be the case when
		\begin{align}\label{eq:spin-wave-equation-of-motion-matrix}
		&\begin{aligned}
			&\det
			\begin{bmatrix}
				(2F_k+1) & (3k^2+q^2)e^{-kL}\\
				e^{-kL} & D(k,q)
			\end{bmatrix}
			=0,
		\end{aligned}
	\end{align}
	with
	\begin{equation}\nonumber
		\begin{aligned}
			D(k,q)=
			(2F_k^*+1)\left[(3k^2+q^2)+2k^3L\cot(qL)/qL\right]&
			\\
			+4k^2(k^2+q^2)\left(1+kL\cot(qL)/qL\right)&.
		\end{aligned}
	\end{equation}
	In the above we used the bulk equation of motion $ 2F_qF_q^*+F_q+F_q^*=0 $ to obtain $ F_q^*/F_q=-(2F_q^*+1) $. Note that we neglected $ 2k^3L\csc(qL)/qL $ in~\cref{eq:exchange-boundary-conditions-constraint-a,eq:spin-wave-equation-of-motion-matrix} since it is exponentially suppressed in the equation of motion and thus not of importance for the dispersion relation. The spin-wave modes hence satisfy
	\begin{equation}
	\begin{aligned}
		&\big[(F_k+1/2)(F_k^*+1/2)-e^{-2kL}/4\big]
		\\\times&(3k^2+q^2)\\
		+&
		(F_k+1/2)2k^2(k^2+q^2)
		\\
		+&
		(F_k+1/2)(F_k^*+1/2+k^2+q^2)\\
		\times&2k^3L
		\cot(qL)/qL\\
		=&0.
	\end{aligned}
	\end{equation}
	When interested in the $ n $-th spin-wave mode the above equation is well approximated by
	\begin{equation}
	\begin{aligned}\label{eq:equation-of-motion-matrix-n-mode-approximate}
			&\big[(F_{k,n}+1/2)(F_k^*+1/2)-e^{-2kL}/4\big]\\
			\times&\big(3k^2+n^2\pi^2/L^2+\delta_{n}\pi^2/4L^2\big)\\
			+&
			(F_{k,n}+1/2)2k^2(k^2+n^2\pi^2/L^2)
			\\
			+&
			(F_{k,n}+1/2)(F_{k,n}^*+1/2+k^2+n^2\pi^2/L^2)
			\\
			\times&2k^3L
			\cot(qL)/qL\\
			=&0,
	\end{aligned}
	\end{equation}
	with $ \delta_{n}\equiv\delta_{n,0} $ the Kronecker delta, $ F_{k,n}\equiv F_k\rvert_{q\rightarrow n\pi/L} $ and $ F^*_{k,n}\equiv F^*_k\rvert_{q\rightarrow n\pi/L} $.
	In order to proceed we use the identity
	\begin{align}
		\pi\cot(\pi x)&=\frac{1}{x}+2x\sum_{n=1}^\infty\frac{1}{x^2-n^2}.
	\end{align}
	To make use of the above identity we consider $ \pi x=q L $. Furthermore, we note that $ q $ for $ n $-th spin-wave is in the interval $ n\pi/L<q<(n+1)\pi/L $. For the $ n $-th spin wave mode we obtain
	\begin{align}\label{eq:cot_approximation}
		\frac{\cot(qL)}{qL}
		\simeq
		\frac{2-\delta_{n}}{q^2L^2-n^2\pi^2}+\frac{2}{q^2L^2-(n+1)^2\pi^2}-\alpha_n,
	\end{align} 
	with $ \delta_n\equiv\delta_{n,0} $ the Kronecker delta and
	\begin{equation}
		\alpha_n=\frac{4}{3\pi^2}\frac{1}{(1+n)^2}.
	\end{equation}
	We expand $ q^2L^2 $ around $ n^2\pi^2 $ for the $ n $-th spin-wave mode. From here it follows that~\cref{eq:equation-of-motion-matrix-n-mode-approximate} can be written explicitly as
	\begin{equation}\label{eq:spin-wave-equation-thirth-order}
		a_kz^3+b_{n,k}z^2+c_{n,k}z+d_{n,k}=0,
	\end{equation}
	with $ z=q^2L^2-n^2\pi^2 $. Furthermore,
	\begin{subequations}
		\begin{align}
			a_{n,k}
			\approx&-1,
			\\
			b_{n,k}=&B_{n,k}-\alpha_nC_{n,k}-a_{n,k}\left(2n+1\right)\pi^2,
			\\
			c_{n,k}=&(4-\delta_n)C_{n,k}
					\\\nonumber
					-&(B_{n,k}-\alpha_nC_{n,k})(2n+1)\pi^2,
					\\
			d_{n,k}=&-C_{n,k}(2-\delta_n)(2n+1)\pi^2,
		\end{align}
	\end{subequations}
	and
	\begin{subequations}\label{eq:spin-qave-dispersion-parameters}
	\begin{align}
	B_{n,k}&\simeq
	\frac{1-e^{-2kL}}{4\gamma_{n,k}/L^2}
	-
	(k^2L^2+n^2\pi^2)
	\\\nonumber
	&\times
	\frac
	{\hskip.5em k^2L^2+n^2\pi^2+\delta_n\pi^2/4}
	{3k^2L^2+n^2\pi^2+\delta_{n}\pi^2/4},
	\\
	2C_{n,k}&\simeq
	\frac
	{k^3L^5\times\gamma_{n,k}^{-1}}
	{3k^2L^2+n^2\pi^2+\delta_{n}\pi^2/4},
	\end{align}
	\end{subequations}
	where $ \gamma_{n,k}=2(\Omega_H+1/2+k^2+n^2\pi^2/L^2) $.
	For the $ n $-th spin-wave mode~\cref{eq:spin-wave-equation-thirth-order} gives the formal solution
\begin{subequations}
		\begin{align}
			z_{n,k}
			&=
			-\frac{1}{3}\frac{b_{n,k}}{a_{n,k}}
			+\sqrt{\frac{-4P_{n,k}}{3}}
			\\\nonumber&\times
			\cos\left[\frac{1}{3}\arccos\left(\frac{3Q_{n,k}}{2P_{n,k}}\sqrt{\frac{-3}{P_{n,k}}}\right)-\frac{2\pi}{3}\right],
	\end{align}
\end{subequations}
	with
	\begin{subequations}
		\begin{align}
			Q_{n,k}
			=&
			\frac{d_{n,k}}{a_{n,k}}
			-\frac{1}{3}\frac{b_{n,k}}{a_{n,k}}\frac{c_{n,k}}{a_{n,k}}
			+\frac{2}{27}\left(\frac{b_{n,k}}{a_{n,k}}\right)^3,
			\\
			P_{n,k}
			=&
			\frac{c_{n,k}}{a_{n,k}}
			-\frac{1}{3}\left(\frac{b_{n,k}}{a_{n,k}}\right)^2.
		\end{align}
	\end{subequations}
	The dispersion relation of the $ n $-th spin-wave mode is accordingly given by~\cref{eq:dispersion_in_terms_of_volume_mode_q}
	\begin{align}\label{eq:spin-wave-dispersion-relation}
		\Omega^2_n&=\left[\Omega_H+\frac{1}{2}+k^2+\frac{n^2\pi^2}{L^2}+\frac{z_n}{L^2}\right]^2-\frac{1}{4}.
	\end{align}
	This is the main result of this paper.
	In~\cref{fig:spin-wave-dispersion-L-24} we compare the analytic dipole-exchange mode in~\cref{eq:spin-wave-dispersion-relation} with the full numeric solution. We see that the analytic dispersion derived above is in good agreement with the full numeric solution of~\cref{eq:exchange-boundary-conditions-plane-wave-ansatz,eq:magnetostatic-boundary-conditions-spin-wave}.
	We like to stress that the analytic spin wave modes given in~\cref{eq:spin-wave-dispersion-relation} do not experience level crossing.
	Hence, the $ n $-th mode does not cross the $ (n-1) $ and $ (n+1) $-th energy modes.
	In the remaining Subsections~\ref{subsec:thick-film-approximation}~and~\ref{subsec:thin-film-approximation} we will simplify~\cref{eq:spin-wave-dispersion-relation} for relative thin and thick films respectively. 
	\subsection{Thick thin-film approximation for the lowest exchange mode}
	\label{subsec:thick-film-approximation}
\begin{figure}
		\includegraphics[width=\columnwidth]{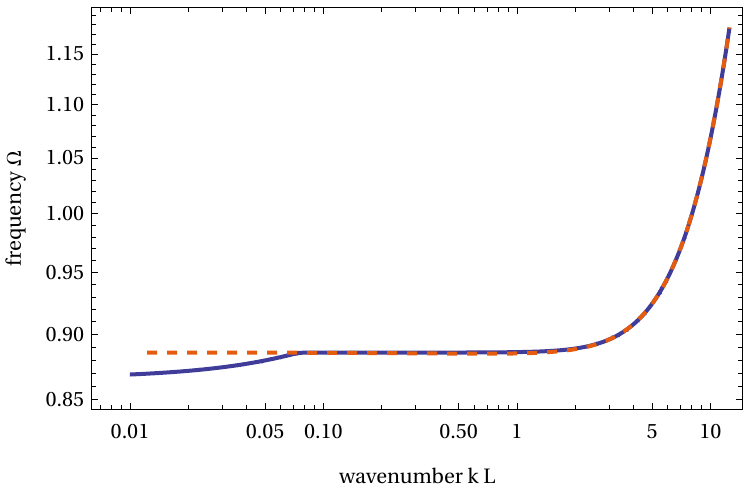}
		\caption{Dipole-exchange dispersion relation of the lowest energy mode for $ \Omega_H=1/2 $ and $ L=24 $. The dashed line correspond to the analytically derived dispersion in~\cref{eq:dispersion-lowest-exchange-mode}, while the solid line gives the numeric solution to~\cref{eq:exchange-boundary-conditions-plane-wave-ansatz,eq:magnetostatic-boundary-conditions-spin-wave} for the lowest energy mode.}
		\label{fig:spin-wave-dispersion-lowest-exchange-mode} 
	\end{figure}

	In relatively thick films, $ L>\mathcal{O}(10\sqrt{J/\mu_0M_s}) $, for sufficiently long wavelengths, the lowest energetic mode will be dominated by the first exchange mode. Away from the DE mode, we may approximate exchange mode solutions of~\cref{eq:spin-wave-equation-of-motion-matrix} by $ D(k,q)=0 $ with $ F^*_k\rightarrow F^*_{k,n} $.
	This is equivalent to $ a_{n,k}\rightarrow0 $ in~\cref{eq:spin-wave-equation-thirth-order} and $ e^{-2k L}\rightarrow0 $ and $ \delta_n\pi^2/4/L^4\rightarrow0 $ in~\cref{eq:spin-qave-dispersion-parameters}. The above results in a second order equation in $ z $ when the exchange mode has a small avoided crossing with the DE mode.
	For the lowest exchange mode $ n\rightarrow0 $ this becomes
	\begin{align}
			2kL\cot(qL)/qL=&4k^2\gamma_{k}-3.
	\end{align}
	Hence, we obtain a quadratic equation in $ z $ 
	\begin{equation}
		b_{k}z^2+c_{k}z+d_{k}=0,
	\end{equation}
	where
	\begin{subequations}
		\begin{align}
			b_{k}=&B_{k}-\frac{4C_{k}}{3\pi^2},
			\\
			c_{k}=&\frac{13}{3}C_{k}-\pi^2B_{k},
			\\
			d_{k}=&-\pi^2C_{k},
		\end{align}
	\end{subequations}
	and
	\begin{subequations}
	\begin{align}
		B_{k}&\simeq3-8(\Omega_H+1/2+k^2)k^2,
		\\
		C_{k}&\simeq2kL.
	\end{align}
	\end{subequations}
	We thus find
	\begin{align}
		z=-\frac{c_k}{2b_k}+\text{sgn}(b_k)\sqrt{\left(\frac{c_k}{2b_k}\right)^2-\frac{d_k}{b_k}}.
	\end{align}
	The lowest exchange mode dispersion is thus
	\begin{align}\label{eq:dispersion-lowest-exchange-mode}
			\Omega^2&=\left[\Omega_H+1/2+k^2+z/L^2\right]^2-1/4.  
	\end{align}
	In~\cref{fig:spin-wave-dispersion-lowest-exchange-mode} we compare the above dispersion relation with the numeric solution of the lowest energy mode. We find good agreement between the approximated dispersion relation in~\cref{eq:dispersion-lowest-exchange-mode} and the numerical lowest energy mode for wavelengths $ k $ larger than the level crossing point with the DE mode.
	Note that the above simplification is not restricted to the lowest energy exchange mode, but can be applied to the higher exchange modes as long as there is no large avoided crossing with the DE mode.
	\subsection{Thin film approximation for the lowest energy mode}
		\begin{figure}[h]
		\includegraphics[width=\columnwidth]{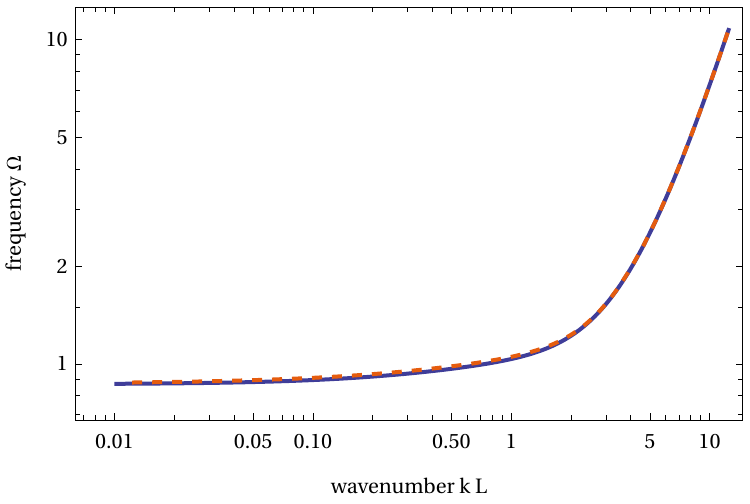}
		\caption{Dipole-exchange dispersion of the lowest energy mode for $ \Omega_H=1/2 $ and $ L=4 $. The dashed line corresponds to the approximate dispersion relation in~\cref{eq:dispersion-lowest-DE-mode}. The solid line corresponds to the numeric solution of~\cref{eq:exchange-boundary-conditions-plane-wave-ansatz,eq:magnetostatic-boundary-conditions-spin-wave}.}
		\label{fig:spin-wave-dispersion-very-thin-film}
	\end{figure}
	\label{subsec:thin-film-approximation}
	For very thin films, $ L\sim\mathcal{O}(\sqrt{J/\mu_0M_s}) $, only the DE wave is of importance for the lowest energy mode. For very thin films it is reasonable to assume $ q^2L^2<(3/4)^2\pi^2 $, we may thus approximate
	\begin{equation}\label{eq:cot_approx_very_thin_film}
		\frac{\cot(qL)}{qL}\simeq\frac{1}{(qL)^2}-\frac{1}{3}.
	\end{equation}
	Thus the lowest energy mode in very thin films is described by a quadratic equation in $ z $
	\begin{equation}
		a_kz^2+b_kz+c_k=0.
	\end{equation}
	Where
	\begin{subequations}
		\begin{align}
			a_k&\approx-2(\Omega_H+1/2+k^2)/L^2,\\
			b_k&\simeq B_k-c_k/3,\\
			c_{k}&\simeq kL/6,
		\end{align}
	\end{subequations}
	and
	\begin{subequations}\label{eq:spin-qave-dispersion-parameters-very-thin-film}
		\begin{align}
			B_{k}&\simeq
			\frac{1-e^{-2kL}}{4}
			-
			\frac{2}{3}(\Omega_H+1/2+k^2)k^2.
		\end{align}
	\end{subequations}
	The lowest energy mode in very thin films is thus given by
		\begin{align}
		z=-\frac{b_k}{2a_k}+\sqrt{\left(\frac{b_k}{2a_k}\right)^2-\frac{c_k}{a_k}},
	\end{align}
	where the lowest energy dispersion relation is given by
	\begin{align}\label{eq:dispersion-lowest-DE-mode}
		\Omega^2&=\left[\Omega_H+1/2+k^2+z/L^2\right]^2-1/4.
	\end{align}
	This is plotted in~\cref{fig:spin-wave-dispersion-very-thin-film}.
	We again find good agreement between the analytic result in~\cref{eq:dispersion-lowest-DE-mode} and the numerical solution of the full boundary conditions in~\cref{eq:exchange-boundary-conditions-plane-wave-ansatz,eq:magnetostatic-boundary-conditions-spin-wave}.
	Note that we took $ \delta_n\pi^2/4L^2\rightarrow0 $ of~\cref{eq:spin-qave-dispersion-parameters} in the very thin-film limit, since it simplifies the expressions for $ B_k $ and $ c_k $ and does not have a big impact on the dispersion relation for very thin-films.

	\subsection{Comparison with Kalinikos and Slavin~\cite{kalinikos_spectrum_1981,kalinikos_theory_1986}}
\begin{figure}[t]
		\includegraphics[width=\columnwidth]{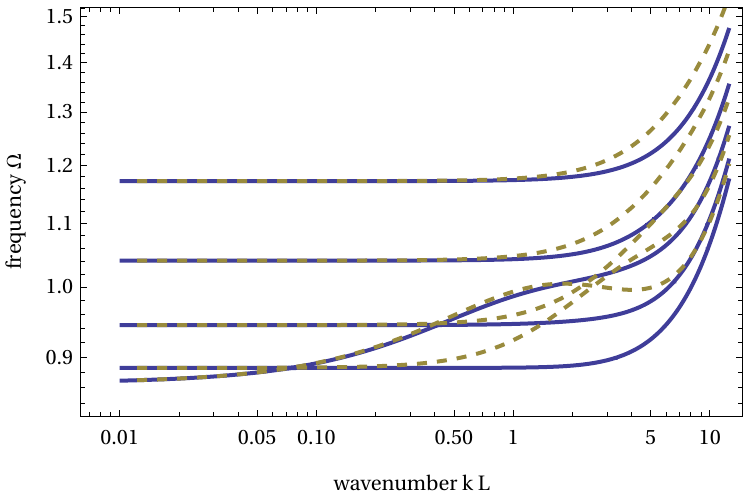}
		\caption{Dipole-exchange spin-wave dispersion relation for $ \Omega_H=1/2 $ and $ L=24 $. The dashed lines correspond to the analytic dispersion derived by Kalinikos and Slavin~\cite{kalinikos_spectrum_1981,kalinikos_theory_1986} in~\cref{eq:spin-wave-dispersion-Kalinikos}. 
		The solid lines correspond to the full numerical solution of~\cref{eq:exchange-boundary-conditions-plane-wave-ansatz,eq:magnetostatic-boundary-conditions-spin-wave}.
		We did not plot the analytic results in~\cref{eq:spin-wave-dispersion-relation,fig:spin-wave-dispersion-L-24}, since they agree very well with the numerical solution. }
		\label{fig:spin-wave-dispersion-L-24-KS}
	\end{figure}

\begin{figure}
		\includegraphics[width=\columnwidth]{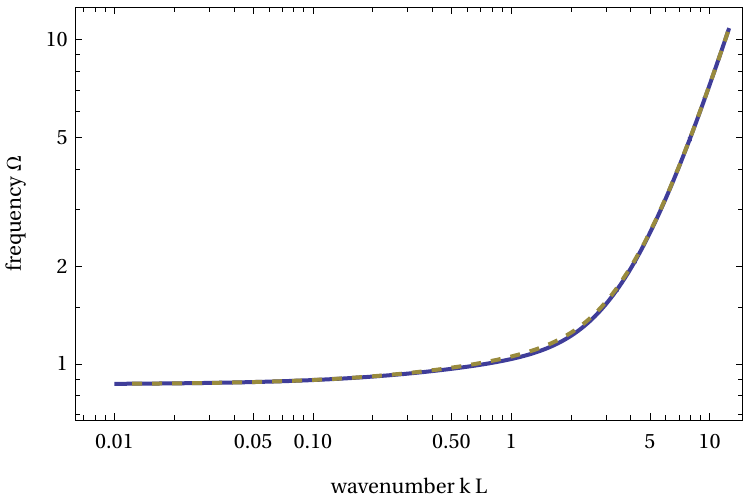}
		\caption{Dipole-exchange dispersion of the lowest energy mode for $ \Omega_H=1/2 $ and $ L=4 $. The dashed line corresponds to the dispersion relation by Kalinikos and Slavin~\cite{kalinikos_spectrum_1981,kalinikos_theory_1986} in~\cref{eq:spin-wave-dispersion-Kalinikos}. The solid line on the other hand corresponds to the numeric solution of~\cref{eq:exchange-boundary-conditions-plane-wave-ansatz,eq:magnetostatic-boundary-conditions-spin-wave}.}
		\label{fig:spin-wave-dispersion-very-thin-film-KS}
	\end{figure}

	In the paper by Kalinikos and Slavin~\cite{kalinikos_spectrum_1981,kalinikos_theory_1986} the dipole-exchange spin-wave spectrum that follows from the diagonal approximation for spin waves propagating perpendicular to a tangentially magnetized thin-film was given by
	\begin{equation}\label{eq:spin-wave-dispersion-Kalinikos}
		\Omega_n^2
		=
		[\Omega_H+1/2+k^2+n^2\pi^2/L^2]^2-[P_n-1/2]^2,
	\end{equation}
	with
		\begin{align}\label{eq:spin-wave-dispersion-Pn-Kalinikos}
		P_n
		=&
		\frac{k^2L^2}{k^2L^2+n^2\pi^2}
		\bigg[
		1
		-\frac{k^2L^2}{k^2L^2+n^2\pi^2}
		\\\nonumber\times&
		\frac{2}{1+\delta_{n}}
		\left(
		\frac{1-(-1)^ne^{-kL}}{k L}		
		\right)
		\bigg].
	\end{align}
	In~\cref{fig:spin-wave-dispersion-L-24-KS} we plotted the spin-wave dispersion by Kalinikos and Slavin~\cite{kalinikos_spectrum_1981,kalinikos_theory_1986} in~\cref{eq:spin-wave-dispersion-Kalinikos} for relatively thick films. We see that the analytic spin-wave dispersion by~Kalinikos and Slavin~\cite{kalinikos_spectrum_1981,kalinikos_theory_1986} shows quantitative differences with the analytic spin-wave dispersion derived in this article~\cref{eq:spin-wave-dispersion-relation,eq:dispersion-lowest-exchange-mode} and the numerical solution of the full problem.
	For very thin films on the other hand, we find good agreement between the theory of~Kalinikos and Slavin~\cite{kalinikos_spectrum_1981,kalinikos_theory_1986}, the analytic results derived in this article~\cref{eq:spin-wave-dispersion-relation,eq:dispersion-lowest-DE-mode} and the numeric solution of the full boundary conditions, see~\cref{fig:spin-wave-dispersion-very-thin-film-KS}.
	\newpage
	\section{Discussion and conclusions}
	We considered the theory of spin waves in ferromagnetic films.
	More specifically, the theory of spin waves propagating perpendicular to an in-plane magnetic field.
	This case  is of special interest since it is the most typical configuration used in spin wave experiments.
	The main result result of this article is the spin wave spectrum in~\cref{eq:spin-wave-dispersion-relation} which we derived by imposed the exchange and magnetostatic boundary conditions on bulk spin wave solutions.
	This derivation differs significantly from the derivation of Kalinikos and Slavin~\cite{kalinikos_spectrum_1981,kalinikos_theory_1986} where the magnetostatic Green's function was used to construct the spin wave spectrum.
	The boundary problem we obtained has an accurate analytical solution which agrees well with the numerical solution and shows quantitative differences with the commonly used theory in~Refs.~\cite{kalinikos_spectrum_1981,kalinikos_theory_1986} in relative thick films.
	This inaccuracy of the spin wave spectrum that follows from the diagonal approximation in the theory by Kalinikos and Slavin~\cite{kalinikos_spectrum_1981,kalinikos_theory_1986}
	has already been observed by~Kreisel\textit{~et al.}~\cite{kreisel_microscopic_2009}.

	Future research could generalize the method to describe spin waves propagating in an arbitrary direction with respect to a generally oriented external magnetic field.
	This is relatively straightforward for in-plane magnetizations. Another way to generalize this model is to include the effects of both surface and boundary anisotropies.
	Lastly, the magnetization profile of spin wave modes could relatively straightforwardly be determined from the spin wave spectrum in~\cref{eq:spin-wave-dispersion-relation,eq:dispersion-lowest-exchange-mode,eq:dispersion-lowest-DE-mode}.
	\section*{acknowledgments}
		J.S.H. would like thank A. R\"uckriegel for helpful discussions. R.A.D. is member of the D-ITP consortium, a program of the Netherlands Organisation for Scientific Research (NWO) that is funded by the Dutch Ministry of Education, Culture and Science (OCW).
		This work is part of the Fluid Spintronics research programme with project
		number 182.069, which is financed by the Dutch
		Research Council (NWO).
	\bibliographystyle{elsarticle-num-names}
	\bibliography{bibliographyspinwavedispersion}
\end{document}